\documentclass{article}

\usepackage{PRIMEarxiv}

\usepackage[utf8]{inputenc} 
\usepackage[T1]{fontenc}    
\usepackage{hyperref}       
\usepackage{url}            
\usepackage{booktabs}       
\usepackage{amsfonts}       
\usepackage{nicefrac}       
\usepackage{microtype}      
\usepackage{lipsum}
\usepackage{fancyhdr}       
\usepackage{graphicx}       
\graphicspath{{media/}}     

\newcommand*\rot{\rotatebox{90}}
\usepackage{subfig}
\usepackage[T1]{fontenc}
\usepackage[utf8]{inputenc}
\usepackage{babel}
\usepackage[font=small,labelfont=bf]{caption}
\usepackage{makecell} 
\usepackage{graphicx}
\usepackage{amsmath}

\usepackage{authblk}

\title{AutoSpeed: A Linked Autoencoder Approach for Pulse-Echo Speed-of-Sound Imaging for Medical Ultrasound}

\author[1,2]{Farnaz Khun Jush}
\author[2]{Markus Biele}
\author[2]{Peter M. Dueppenbecker}
\author[1]{Andreas Maier}
\affil[1]{Pattern Recognition Lab, Friedrich-Alexander-University, Erlangen, Germany}
\affil[2]{Technology Excellence, Siemens Healthcare GmbH, Erlangen, Germany}
\affil[ ]{\textit {\{farnaz.khun.jush,andreas.maier\}@fau.de}}
\affil[ ]{\textit {\{markus.biele,peter.dueppenbecker\}@siemens-healthineers.com}}

\begin{document}
\maketitle

\begin{abstract}
Quantitative ultrasound, e.g., speed-of-sound (SoS) in tissues, provides information about tissue properties that have diagnostic value. 
Recent studies showed the possibility of extracting SoS information from pulse-echo ultrasound raw data (a.k.a. RF data) using deep neural networks that are fully trained on simulated data.
These methods take sensor domain data, i.e., RF data, as input and train a network in an end-to-end fashion to learn the implicit mapping between the RF data domain and SoS domain. However, such networks are prone to overfitting to simulated data which results in poor performance and instability when tested on measured data. 
We propose a novel method for SoS mapping employing learned representations from two linked autoencoders. We test our approach on simulated and measured data acquired from human breast mimicking phantoms. We show that SoS mapping is possible using linked autoencoders. The proposed method has a Mean Absolute Percentage Error (MAPE) of $2.39\%$ on the simulated data.
On the measured data, the predictions of the proposed method are close to the expected values with MAPE of $1.1$~$\%$.
Compared to an end-to-end trained network, the proposed method shows higher stability and reproducibility. 

\end{abstract}


\section{Introduction}
\label{sec1}

B-mode imaging is a qualitative approach and its outcome is highly dependent on the operator's expertise. 
Quantitative values, e.g., speed-of-sound (SoS) in tissue, can ease interpretation and provide additional information about tissue properties. In particular, SoS is proven to carry diagnostic information for differentiating tissue types in breast cancer screening \cite{sak2017using,sanabria2018breast,schreiman1984ultrasound}.   

Prior works attempted to reconstruct SoS from ultrasound pulse-echo data. For instance,
\cite{sanabria2018speed} used a passive reflector to record the time of flight of the signals and reconstructed SoS by solving a limited angle inverse problem. 
Although a passive reflector provides an acoustic mirror effect and a timing reference, placing such a reflector is not feasible in many clinical setups.  
\cite{stahli2020improved} used spatial phase shifts of beamformed echoes obtained under varying steering angles.
One of the problems of this approach is that small variations in the inputs, for instance in the presence of measurement or phase noises can make the setup unstable \cite{stahli2020bayesian}. 
Thus, \cite{stahli2020bayesian} included an a priori model based on B-mode image segmentation in a Bayesian framework. 
This model relies on manual segmentation of the B-mode images which might not be practical in clinical setups. 
In the pulse-echo setup, model-based SoS reconstruction is non-trivial due to limitations in the number of transmission angles and limited frequency bandwidths, thus, SoS reconstruction by analytical or optimization approaches requires carefully chosen regularization and optimization methods, prior knowledge, and complex fine-tuning \cite{vishnevskiy2018image}.

During data acquisition, the sensor encodes an intermediate representation of the object under examination in the sensor domain. 
The reconstruction then is performed by inversion of the corresponding encoding function \cite{zhu2018image}. 
However, the exact inverse function is not available a priori, thus, reconstruction problems need to approximate the function. 
Recently, deep neural networks are vastly being used to solve reconstruction problems. 
Opposed to analytical and optimization methods where the inverse function is approximated in multiple stages of signal processing and/or optimization steps, in deep learning approaches, the network tries to solve the problem by learning the corresponding mapping between the sensor domain and reconstruction domain \cite{zhu2018image}. 
Therefore, during the training, a low-dimensional representation of the data in both domains is implicitly learned. 
This concept was first proposed in \cite{zhu2018image}, a.k.a. AUTOMAP, to perform a robust reconstruction of multiple sensor-domain data. 
The idea is derived from domain transfer approaches, yet, no mechanism is implemented to actually guarantee the robustness of the solutions found by such black-box networks. 
Consequently, these methods are proved to be unstable, especially when the test data has perturbations or structural differences compared to training data \cite{antun2020instabilities}. 

Similar techniques are employed for SoS reconstruction from ultrasound echo data. 
\cite{feigin2019deep,feigin2020detecting,heller2021deep,jush2020dnn,khun2021data,jush2022deep,oh2021neural} investigated encoder-decoder networks with multiple or single steering angles for SoS reconstruction. 
In these studies, an encoder-decoder network takes sensor domain data as input and directly reconstructs the SoS map in the output. 
Since there is no known gold standard method to create SoS GT from measured pulse-echo ultrasound data, all the investigated methods rely on simulated data. 
As such, transfer of these methods to clinical setups is challenging and their robustness is still under debate \cite{feigin2019deep,jush2022deep}. 
These methods are prone to overfitting the distribution of the simulated data and thus perform unsatisfactorily when tested on real data \cite{jush2022deep}.

\begin{figure}[!htbp]
\subfloat[]{\includegraphics[scale=0.75]{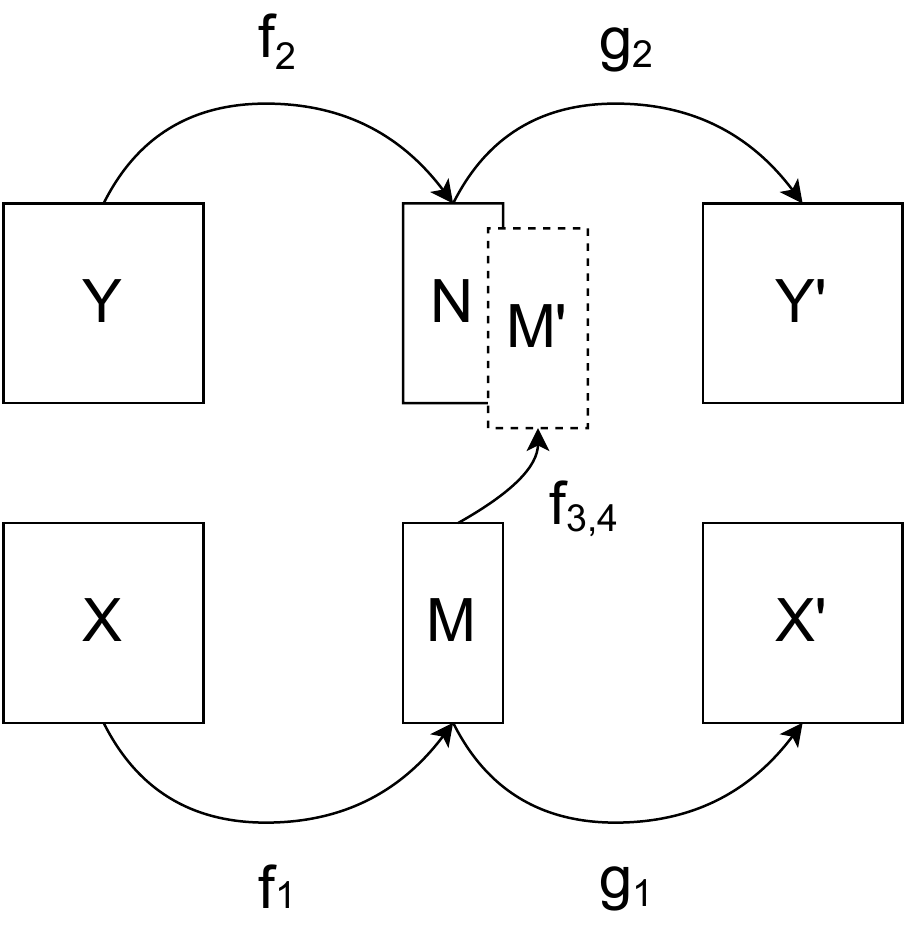}}\hfill
\subfloat[]{\includegraphics[scale=0.75]{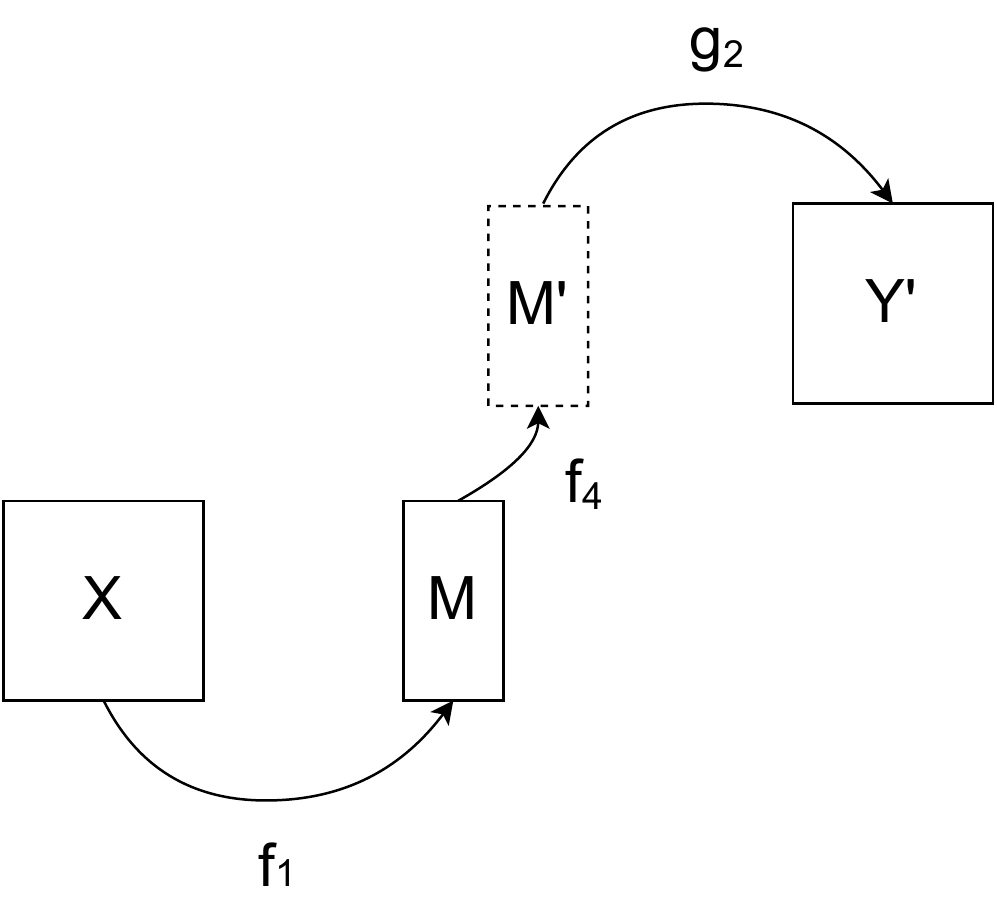}}
\caption{Overview of the proposed method: $X$ being RF domain, $Y$ being SoS domain, $f$ encoder and $g$ decoder \newline 
(a) Training: firstly, two domain are encoded to an intermediate low-dimensional representation by training two linked autoencoders jointly: $X \rightarrow f_1(X) = M \rightarrow g_1(f_1(X)) = X' \approx X $ and $Y \rightarrow f_2(Y) = N \rightarrow g_2(f_2(Y))  = Y' \approx Y $ ; Secondly, a mapping is performed between two representations during the training ($M \rightarrow f_3(M) = M' \approx N $) and in an additional fine-tuning step ($M \rightarrow f_4(M) = M' \approx N$); \newline 
(b) Inference: without further training, two steps are joined to create a network that maps RF domain to SoS domain: $X \rightarrow f_1(X)=M \rightarrow f_4(f_1(X)) = M' \rightarrow g_2(f_4(f_1(X)))=Y'$.} 
\label{fig:overview}
\end{figure}

In this study, we propose a novel approach for SoS mapping from a single plane-wave acquisition.
Instead of the end-to-end encoder-decoder approaches previously proposed, where an implicit mapping between sensor domain and SoS domain was learned, following the known operator paradigm \cite{maier2019learning}, we propose a method to break the problem into two steps:
In the training phase, firstly, we encode the sensor domain and the SoS domain data in an intermediate low-dimensional representation using two linked autoencoders; secondly, we find a mapping between two representations. 
For inference, without any further training, these two steps are joined to create a network that takes the RF data as input and returns the SoS map in the output. An overview is demonstrated in Fig. \ref{fig:overview}. 
We then compare the proposed method with a baseline encoder-decoder network both on simulated and measured data acquired from human breast mimicking phantom.

\section{Methods}

\subsection{Network Architecture, Training}

Autoencoders \cite{bengio2009learning,bengio2013representation,maier2019gentle} are vastly being used to learn low-dimensional latent representations. 
An autoencoder is a neural network that maps its input to itself utilizing an intermediate representation, the so-called latent space. Autoencoders thus learn to map input data to the latent space, and latent space data back to the input domain \cite{chen2017deep,masci2011stacked,vincent2008extracting}. 
Hence, to learn the efficient representation from two domains, i.e., sensor domain (representing RF data) and SoS domain, autoencoders can be employed.
Convolutional autoencoders combine local convolution connections with autoencoders, thus, preserving spatial locality \cite{chen2017deep}. 

As such, to extract an efficient representation from two domains, firstly two linked convolutional autoencoders are trained jointly.
Secondly, the encoded latent spaces are extracted and a final step of training is performed to fine-tune the mapping of RF data and SoS map latent spaces. 

\subsubsection{Linked Autoencoders}

\begin{figure}[!htbp]
\centering
\subfloat[]{\includegraphics[scale=0.5]{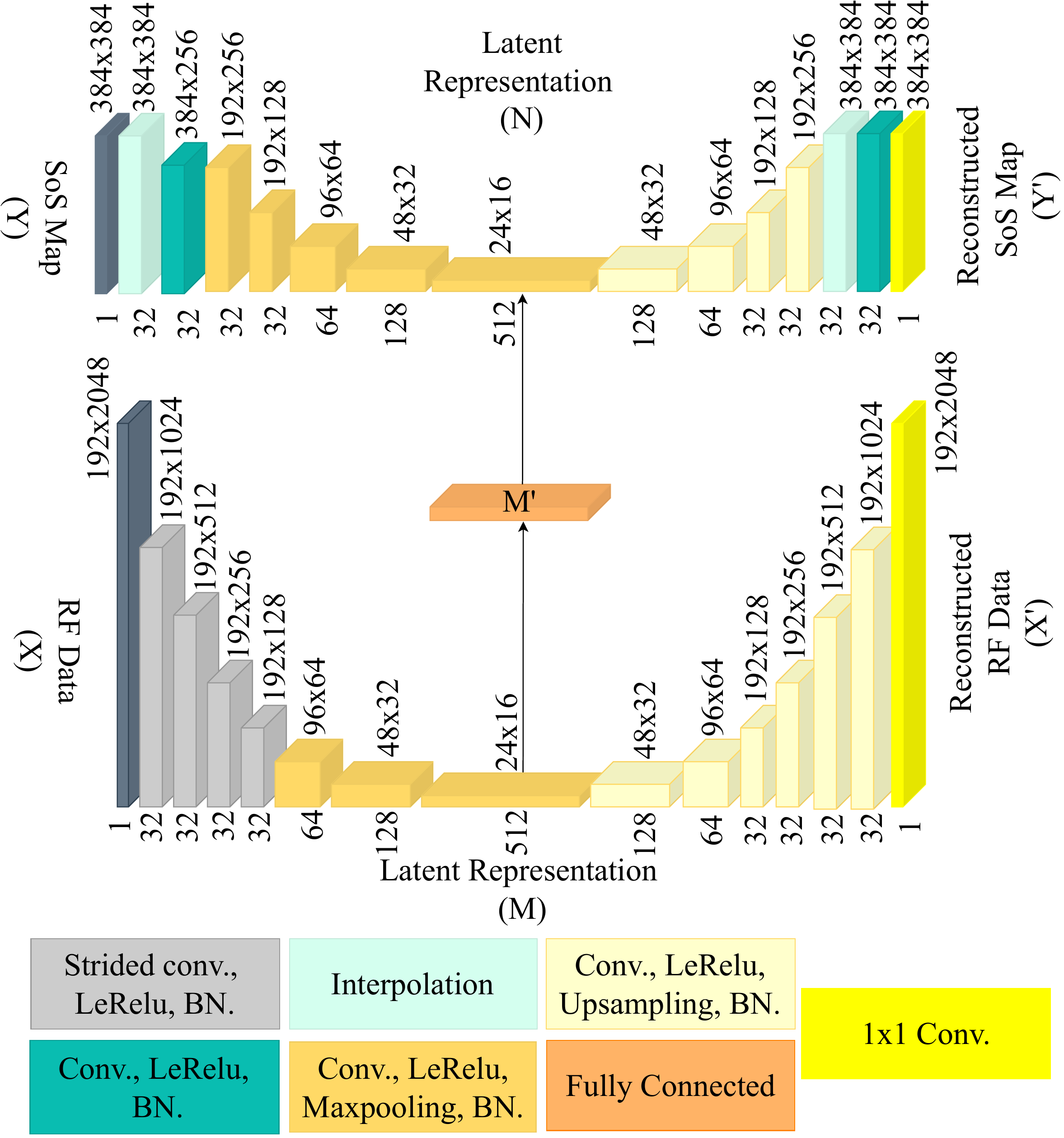}}\hspace{0.5mm} 
\\
\centering
\subfloat[]{\includegraphics[scale=0.5]{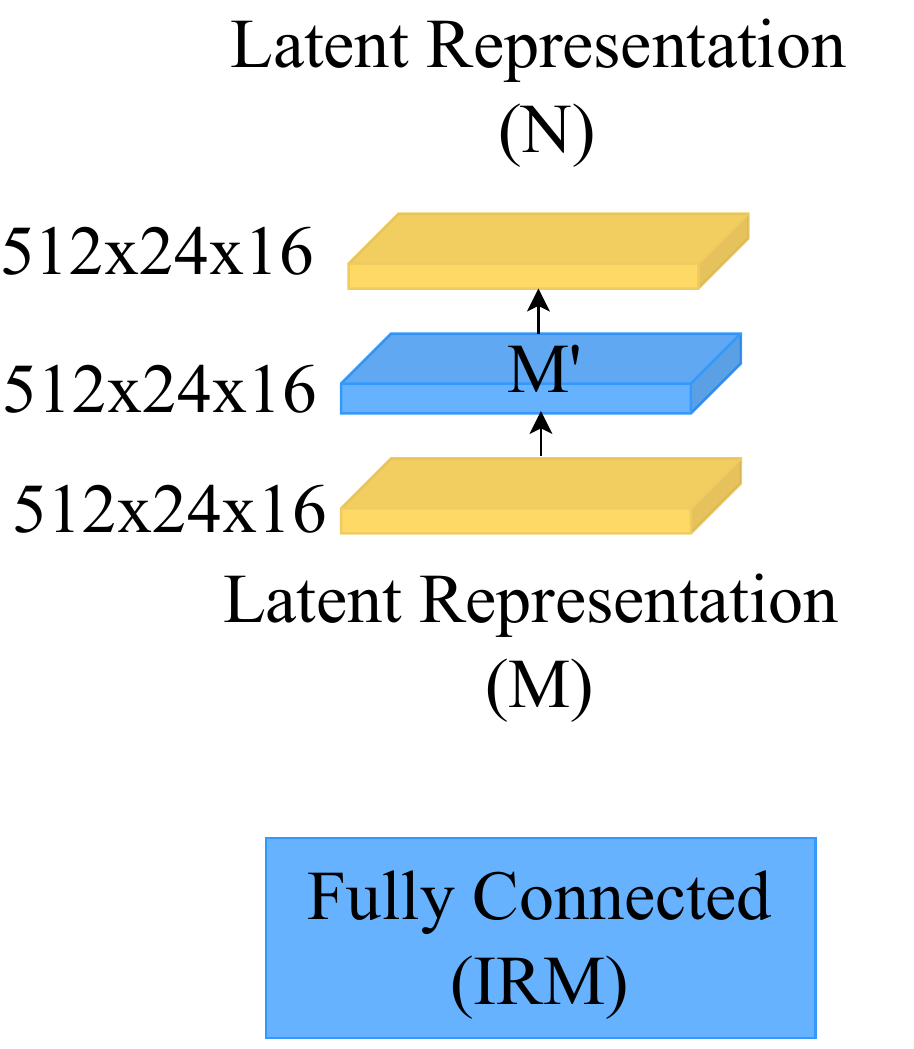}}\hspace{0.5mm}\\
\caption{\textbf{Network Architecture, Training:} (a): Linked Autoencoders: two convolutional autoencoders connected via their latent spaces, the linked autoencoders are trained jointly to encode two domains where the intermediate representation match as closely as possible. One autoencoder takes the SoS map as input and the other takes RF data as input. The RF data is a matrix of size $192 \times 2048$ ($192$ channels, $2048$ samples with a sampling frequency of $40$~$MHz$). 
The SoS map is a matrix of size $384 \times 384$ which translates to $3.84$~$cm$ in lateral and axial directions. 
Operations in each layer are color-coded. 
(b): IRM-Layer: the intermediate representation from the RF domain is mapped to the intermediate representation from the SoS domain via a fully connected layer. The color blocks show the operations in each layer. The network is implemented using Tensorflow 2.2 \cite{tensorflow2015-whitepaper}.   }
\label{fig: training}
\end{figure}

Two convolutional autoencoders that are connected via their latent spaces are trained to extract robust features of RF data and SoS map.
During training efficient representations of the SoS domain and RF domain are encoded in the latent spaces of the autoencoders. However, these representations can have different interpretations. To bridge the gap between the aforementioned representations, they are connected via a fully connected layer that is trained simultaneously with two autoencoders. 
This way, the autoencoders are optimized in a manner that the latent spaces of two networks indicate the closest possible representations. Fig.~\ref{fig: training} (a) demonstrates the model architecture. %

The autoencoders take the input vectors of RF and SoS domain as $x$ and $y$, respectively, and map them to the hidden representations $m$ and $n$, where $ m=f_{\theta_1}(x)=\xi(x\ast W_1+b_1)$ and $n=f_{\theta_2}(y)=\xi(y\ast W_2+b_2)$ are parameterized by $\theta_1=\{W_1,b_1\}$ and $\theta_2=\{W_2,b_2\}$, respectively. 
$W_{1,2}=\{w_j,j=1,2,..,k\}$, where $w_j$ is the weight of convolution kernel $j$ and $k$ is the kernel size \cite{bengio2013representation,masci2011stacked}. $b_1$ and $b_2$ are biases. $f_{\theta}$ is the encoder. $\xi$ is the activation function \cite{xu2015empirical}. 

The encoded representations, $m$ and $n$, are then used to reconstruct $x'$ and $y'$, where $x'=g_{\theta'_1}(m)=\xi(m\ast W'_1+b'_1)$ and $ y'=g_{\theta'_2}(n)=\xi(n\ast W'_2+b'_2)$ are similarly parametrized by $\theta'_1=\{W'_1,b'_1\}$ and $\theta'_2=\{W'_2, b'_2\}$, respectively. $b'_1$ and $b'_2$ are biases. $W'_{1,2}=\{w'_j, j=1,2,..,k\}$, where $w'_j$ is the weight of convolution kernel $j$ and $k$ is the kernel size \cite{bengio2013representation,masci2011stacked}. $g_{\theta'}$ is the decoder.

Each training $x^{(i)}$ and $y^{(i)}$ is thus mapped to a corresponding $m^{(i)}$ and $n^{(i)}$ and reconstructs $x'^{(i)}$ and $y'^{(i)}$. 

Simultaneously, the latent spaces are being optimized to match: $m$ is mapped to $m'$ via a fully connected layer, $m'=f_{\theta_3}(m)=\xi(m W_3 + b_3)$, $\theta_3$ is parameterized by $W_3$, the weight, and $b_3$, the bias. During the training $m', n$ are being optimized in the loss function to match closely, where $m'=f_{\theta_3}(f_{\theta_1}(x))$ and $n = f_{\theta_2}(y)$. The parameters of the model are optimized to minimize the following cost function: 

\begin{equation}
\begin{aligned}
    J_{LAE} =  \frac{1}{n} \sum_{i=1}^{N} L(x^{(i)},x'^{(i)}) + L(y^{(i)},y'^{(i)}) + L(m'^{(i)},n^{(i)})\\ 
    = \frac{1}{n} \sum_{i=1}^{N}  L(x^{(i)},g_{\theta'_1}(f_{\theta_1}(x^{(i)})) 
    + L(y^{(i)},g_{\theta'_2}(f_{\theta_2}(y^{(i)})) \\
    +  L(f_{\theta_3}((f_{\theta_1}(x^{(i)})),f_{\theta_2}(y^{(i)}))
\end{aligned}
\end{equation}

where L is the loss function, here the Mean Squared Error (MSE), $L(x,x')=~||x-x'||^2$, is used. Note that for simplification, vector representation is used in the text but Fig.~\ref{fig: training} shows matrix representation.

\subsubsection{Fine-tuning (IRM-Layer)} 
After convergence of the linked autoencoders, the latent spaces of two autoencoders are extracted. 
To fine-tune the mapping of the latent spaces, a fully connected layer which will be referred to as the Intermediate Representation Mapper Layer (IRM-Layer) is trained separately. This layer takes the latent space of the RF autoencoder as input (\(n\)) and the latent space of the SoS autoencoder as output (\(m\)) similar to the mapping of latent spaces during linked autoencoder training.

Thus, $m$ is mapped to $m'$, where $m'=f_{\theta_4}(m)=\xi (m W_4 + b_4)$. $\theta_4$ is parameterized by $W_4$, the weight and $b_4$ the bias. 
The following cost function is optimized: 
\begin{equation}
    J_{IRM} = \frac{1}{n}\sum_{i=1}^{N}L(n,m') = \frac{1}{n}\sum_{i=1}^{N}L(n,f_{\theta_4}(m)) 
\end{equation}

IRM-Layer represents a mapping function between two latent spaces \(m\) and \(n\). Fig.~\ref{fig: training} (b) shows the schematic.

\subsection{Network Architecture, Inference}

For the inference, we propose a network henceforth will be referred to as AutoSpeed. 
AutoSpeed is an encoder-decoder network that its encoder is detached from the trained RF autoencoder and its decoder is detached from the trained SoS autoencoder. 
Extracted encoder and decoder paths are connected via the trained IRM-Layer. 
Note that no retraining step is performed to employ the AutoSpeed, the layers are simply taken from the trained linked networks and connected via the trained IRM-Layer. 

In the encoder, $x$ is mapped to $m$, where $m =f_{\theta_1}(x)=\xi(x*W_1+b_1)$. $b_1$ and $W_1$ are bias and weights of the encoder of the RF autoencoder. The IRM-Layer with weights and bias $W_4$ and $b_4$ is then being used to map $m$ to $m'$: $m'=f_{\theta_4}(m)=\xi (m W_4 + b_4)$. 

\begin{figure}[!htbp]
\centering
\includegraphics[scale=0.5]{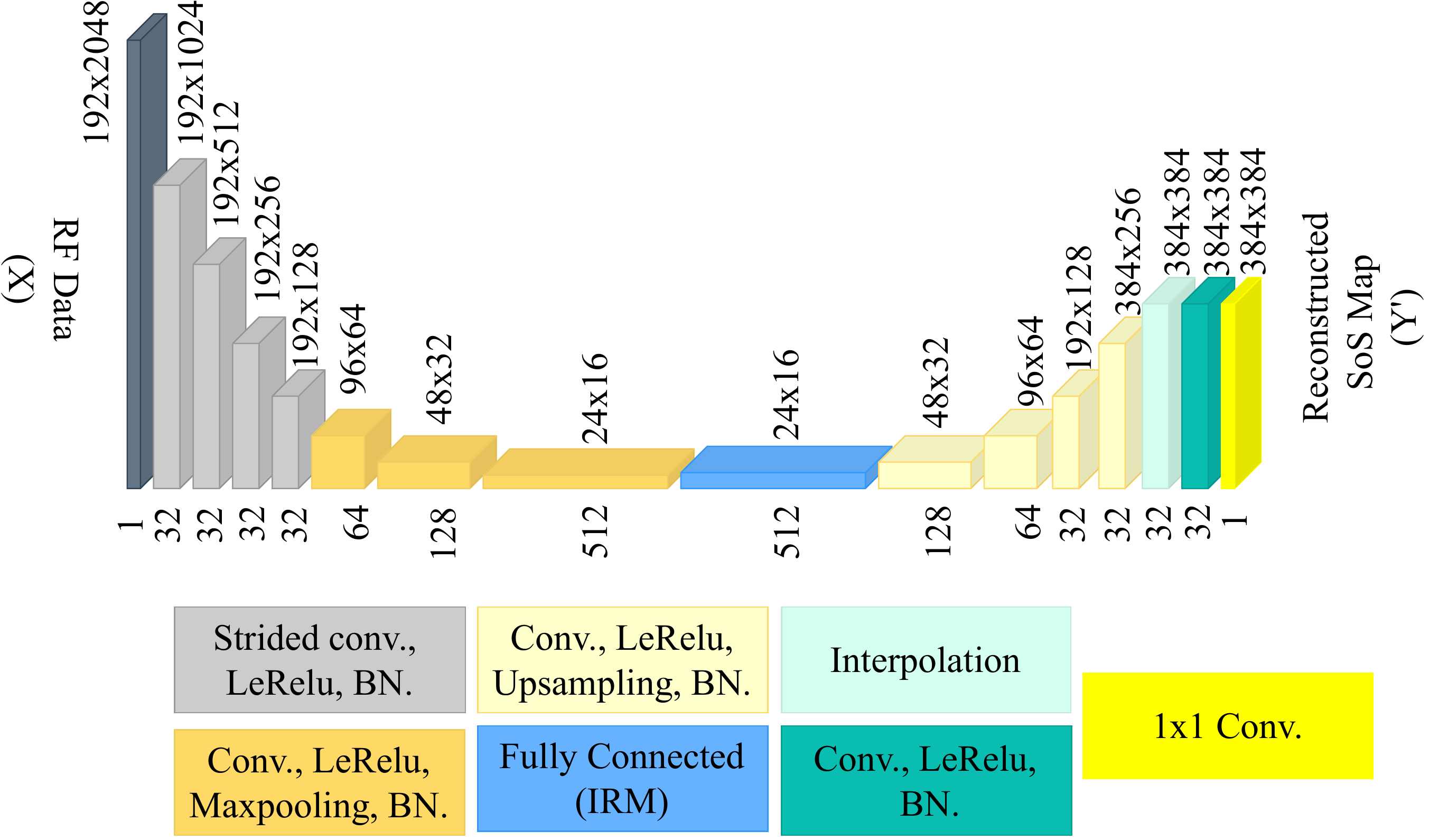}
\caption{\textbf{Network Architecture, Inference (AutoSpeed)}: The network consists of an encoding path that is extracted from the encoding path of the trained RF Autoencoder and a decoding path that is extracted from the trained SoS Autoencoder, two paths are connected via a fully connected layer, i.e., IRM-Layer, which is trained during the fine-tuning step. No further training is required to employ AutoSpeed.
AutoSpeed takes RF data as input and returns the SoS map in the output.} 
\label{fig: AutoSpeed}
\end{figure}

In the decoder, $m'$ is mapped to $y'= g_{\theta'_{2}}(m')=\xi(m'*W_2+b_2)$, where $W_2$ and $b_2$ are the weights and biases are taken from decoder of the SoS autoencoder. Fig.~\ref{fig: AutoSpeed} shows the schematic of AutoSpeed. 

\section{Results}

\subsection{Dataset}

For SoS reconstruction, acquiring sufficient measured data alongside its corresponding GT for training a deep neural network is challenging because there is no known gold-standard method capable of creating exact GT for reflection data (pulse-echo ultrasound) and there are only a few phantoms available with known heterogeneous SoS. 
Therefore, for deep learning-based approaches using simulated data for training is a common practice \cite{vishnevskiy2018image,feigin2019deep,feigin2020detecting,heller2021deep,jush2020dnn,khun2021data,jush2022deep,oh2021neural}. 
K-Wave toolbox \cite{treeby2010k} (Version 1.3) is used for the simulation of training data.
K-Wave is a MATLAB toolbox that allows the simulation of linear and non-linear wave propagation with arbitrary heterogeneous parameters \cite{treeby2010k}.

The LightABVS system that has been previously introduced for pulse-echo ultrasound imaging applications is used for data acquisition \cite{hager2019lightabvs}. LightABVS system consists of a linear probe with 192 active channels, a pitch of $200~\mu m$, and a center frequency of $5$~$MHz$.
The size of the medium is considered to be $3.8~cm$ in depth and $7.6~cm$ in the lateral direction (twice the size of the probe head) and the probe head is placed above the central section of the medium. A single plane-wave with zero degrees is used for the simulations \cite{feigin2020detecting,jush2020dnn,jush2022deep}. 
The medium is simulated on a 2D grid of size \(1536\times3072\). 

The simulation setup is based on the Combined setup proposed by \cite{jush2022deep}, where a joint set of datasets including both geometrical and natural tissue models are simulated for training data generation:
\begin{itemize}

    \item Ellipsoids: The geometrical mediums are based on a simplified model for organs proposed by \cite{feigin2019deep} and adopted for breast use-case in \cite{jush2020dnn,jush2022deep}. This setup considers a medium with homogeneous background and embeds multiple ellipsoids in random places inside the homogeneous background. Random SoS values in the range $[1300-1700]~m/s$ are assigned to the medium structures in the SoS domain. 
    
    \item T2US: The natural tissue structures are extracted from Tomosynthesis images. This setup is based on the T2US setup proposed in \cite{jush2022deep}. 
    Random patches from Tomosynthesis images with benign or malignant lesions are used to create SoS maps with SoS values in the range $[1300-1700]~m/s$. 

\end{itemize}
 
The attenuation value is set to $0.75$~$dB/MHz.cm$ and the mass density of $1020$~$kg/m^3$ based on the property of breast tissue \cite{szabo2004diagnostic}, Table B.1. 
Further details of this setup can be found in \cite{jush2022deep}.

\subsection{Training}
\subsubsection{Linked Autoencoder}
The linked autoencoder is trained for 200 epochs with SGD optimizer with a batch size of $8$ on an Nvidia RTX Quadro 8000 GPU. 
During training, $6000$ training cases were used ($4800$ training, $1200$ validation). 
The test set contains 150 cases. The size of the dataset is chosen based on \cite{feigin2019deep,jush2020dnn,jush2022deep}. 

\textbf{RF Autoencoder:} Over 10 training runs, the RF autoencoder converges to an average Root Mean Squared Error (RMSE) of $6.06 \pm 0.31$ (SoS$\pm$SD where SD is the standard deviation) and Mean Absolute Error (MAE) of $3.94 \pm 0.29$, where the input RF data is in the range $[-1024,1024]$. 
It is noteworthy that during training experiments we observed that normalization hinders the convergence of RF Autoencoder, thus, the data is not normalized.

\textbf{SoS Autoencoder:} The SoS autoencoder converges to RMSE of $7.89\pm 3.23 $~$m/s$ and MAE of $6.68 \pm 3.03 $~$m/s$, where the input range is $[1300,1700]$~$m/s$. 

\textbf{Intermediate Layer:} The intermediate representation converges to RMSE of $1.14 \pm 0.28$ and MAE of $0.93 \pm 0.36$, where the output of mid-layers is in the range $[-15,30]$. 

\subsubsection{Fine-tuning (IRM-Layer)}
The IRM-layer is trained for 150 epochs using the Adam optimizer \cite{kingma2014adam}, where it converges to RMSE of $0.68 \pm 0.28$ and MAE of $0.55 \pm 0.19$ over 10 training runs.

\subsection{Inference}
\subsubsection{Simulated data} 

On the test dataset, AutoSpeed has an average RMSE of $47.98 \pm 4.15$~$m/s$, MAE of $37.26 \pm 3.56$~$m/s$ and and MAPE of $2.39 \pm 0.22 \%$ over 10 training runs. 

We set up a network based on \cite{jush2022deep} and compared the results of AutoSpeed with this network. 
Hereafter, we will refer to the baseline network as En-De-Net. The En-De-Net converges to RMSE of $23.64 \pm 0.7$~$m/s$, MAE $17.32 \pm 0.7$~$m/s$, and MAPE of $1.11 \pm 0.05\%$. 
The error ranges of both networks are in the typical range of SoS errors $(1-2\%)$ \cite{hill2004physical,oh2021neural,feigin2019deep,jush2022deep}. 
However, on the simulated data the En-De-Net has a lower error rate.

\subsubsection{Measured data} 

\textbf{Single Frames:} We acquired ultrasound RF data for a CIRS multi-modality breast phantom with the LightABVS system \cite{hager2019lightabvs}. 
$30$ cases with visible dense masses are acquired, where masses are placed in different locations inside the region of interest (ROI).  
The phantom contains an anechoic skin-mimicking layer and hyperechoic dense masses embedded in homogeneous background. Because we lack measured SoS ground truth data, we compare to the manufacturer's specifications. The predicted SoS in the background region and dense lesions region is in the range specified by the manufacturer in Table~\ref{table: errors}, Expected value.
Each inclusion is manually segmented based on the B-mode image to compare SoS values predicted inside the inclusion and in the background by two networks.
Since the simulation model does not include anechoic tissue characteristics, the skin-mimicking layer is excluded in the error computation in Table~\ref{table: errors}.

\begin{figure}[!t]
    \begin{minipage}[t]{0.99\textwidth}
        \centering
        \includegraphics[scale=0.35]{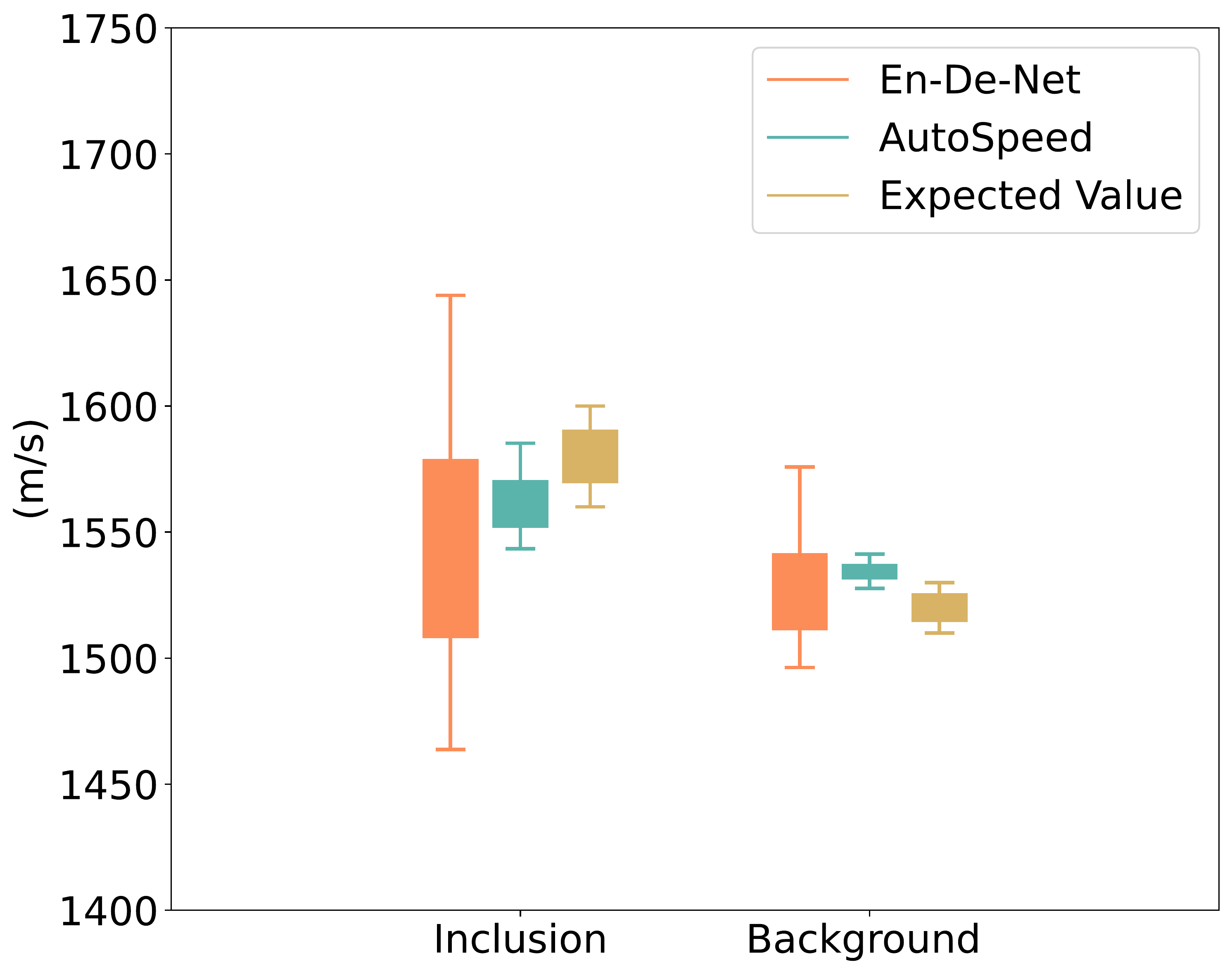}
        \captionof{figure}{Predicted SoS inside dense inclusion and in the background for 30 measured phantom cases.}
         \label{fig: boxplot}
    \end{minipage}
\end{figure}

\begin{figure*}[!t] 
\centering 

\renewcommand{\arraystretch}{0.05}
\begin{tabular}{@{\hspace{0.5mm}} c @{\hspace{0.5mm}}c @{\hspace{0.5mm}}c @{\hspace{0.5mm}}c @ {\hspace{0.5mm}}c @{\hspace{0.5mm}}c  @{\hspace{0.5mm}}c  @{\hspace{0.5mm}}c  @{\hspace{0.5mm}}c @{\hspace{0.5mm}}c @{\hspace{0.5mm}}c}
& {  \#1} & { \#2} & { \#3} & { \#4} & { \#5}  & {  \#6} & { \#7} & { \#8} & &\\

\raisebox{-.5\totalheight}{  \rot{B-mode}} & 
\raisebox{-.5\totalheight}{\includegraphics[  width =1.9cm]{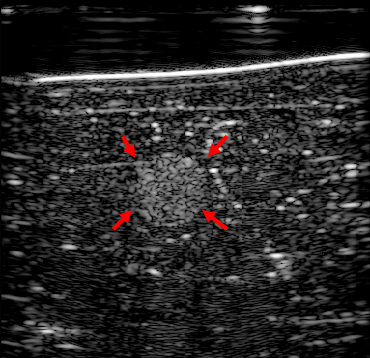}}&
\raisebox{-.5\totalheight}{\includegraphics[  width =1.9cm]{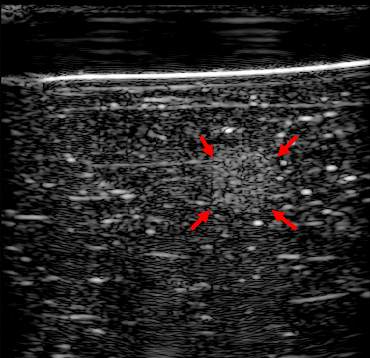}}&
\raisebox{-.5\totalheight}{\includegraphics[  width =1.9cm]{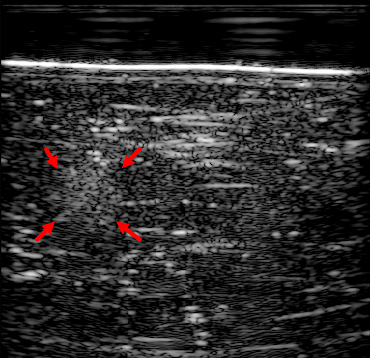}}&
\raisebox{-.5\totalheight}{\includegraphics[  width =1.9cm]{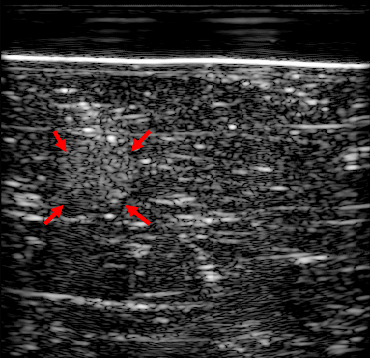}}&
\raisebox{-.5\totalheight}{\includegraphics[  width =1.9cm]{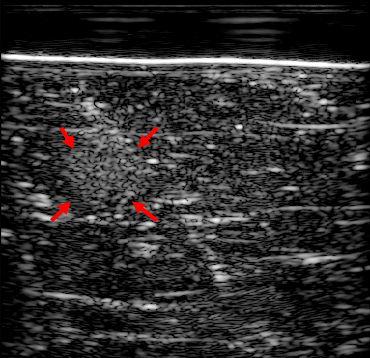}}&
\raisebox{-.5\totalheight}{\includegraphics[  width =1.9cm]{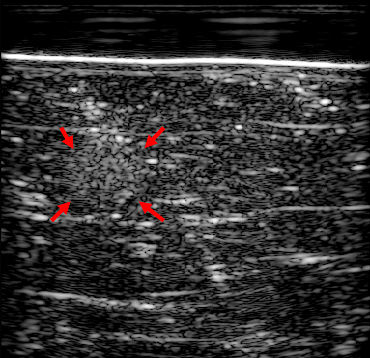}}&
\raisebox{-.5\totalheight}{\includegraphics[  width =1.9cm]{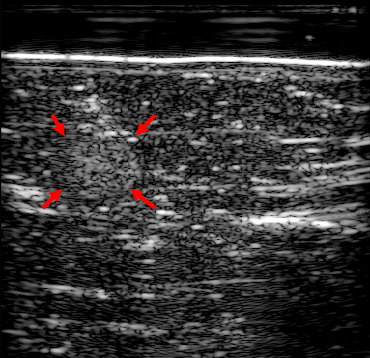}}&
\raisebox{-.5\totalheight}{\includegraphics[  width =1.9cm]{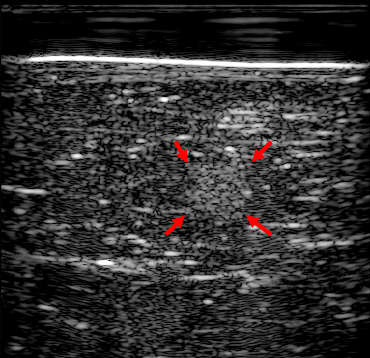}}&
\raisebox{-.5\totalheight}{\includegraphics[height =1.9cm]{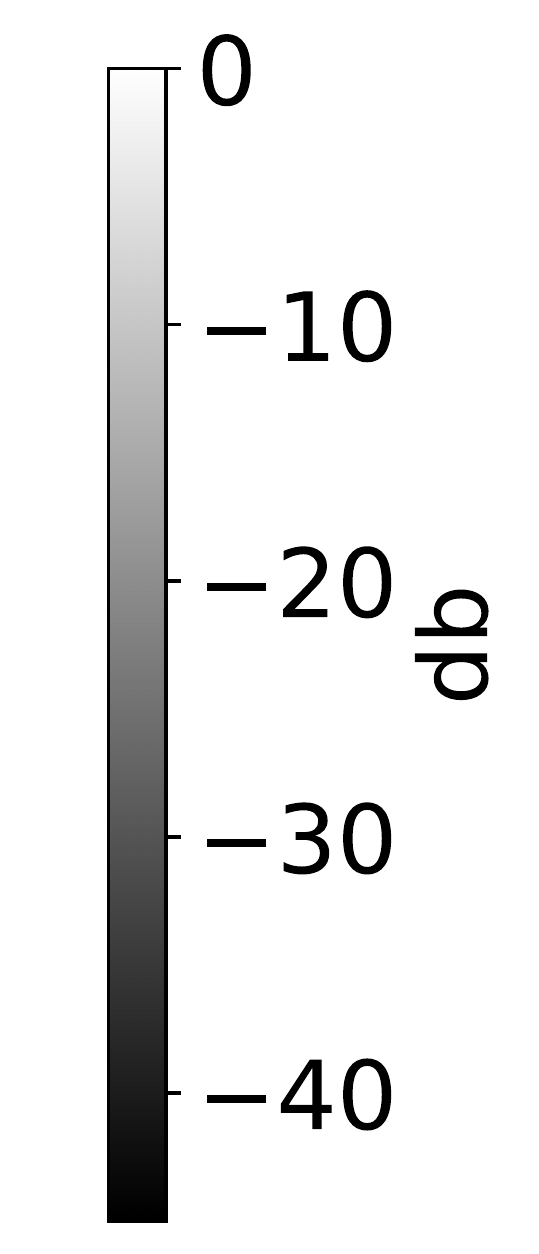}}\\

 \raisebox{-.5\totalheight}{ \small \rot{\makecell{{En-De-Net} \\ {{SoS Map}} }}} & 
 \raisebox{-.5\totalheight}{\includegraphics[  width =1.9cm]{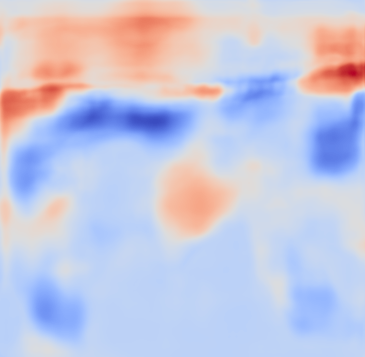}}&
 \raisebox{-.5\totalheight}{\includegraphics[  width =1.9cm]{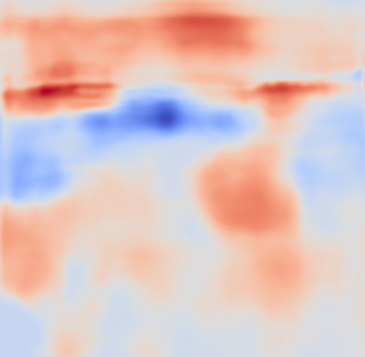}}&
 \raisebox{-.5\totalheight}{\includegraphics[  width =1.9cm]{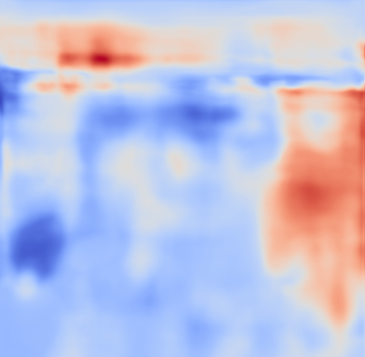}}&
 \raisebox{-.5\totalheight}{\includegraphics[  width =1.9cm]{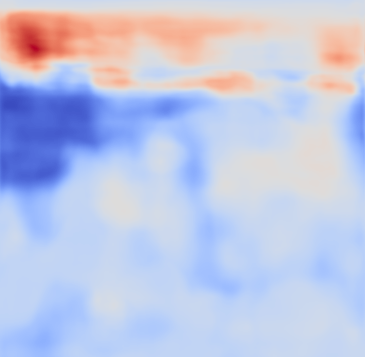}}&
 \raisebox{-.5\totalheight}{\includegraphics[  width =1.9cm]{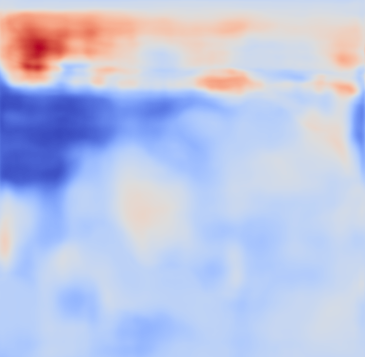}}&
 \raisebox{-.5\totalheight}{\includegraphics[  width =1.9cm]{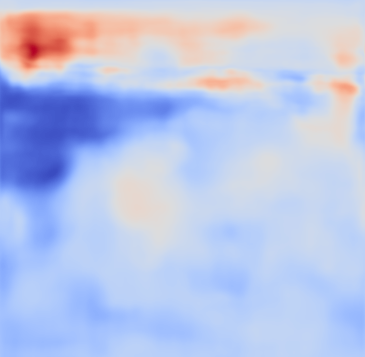}}&
 \raisebox{-.5\totalheight}{\includegraphics[  width =1.9cm]{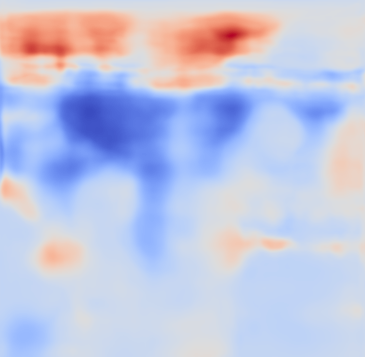}}&
 \raisebox{-.5\totalheight}{\includegraphics[  width =1.9cm]{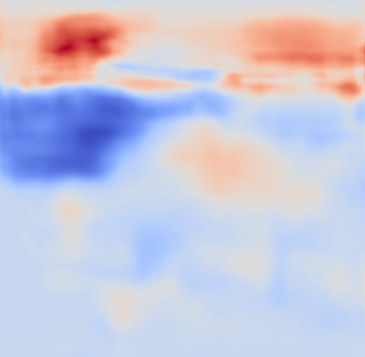}}&
\raisebox{-.5\totalheight}{\includegraphics[height =1.9cm]{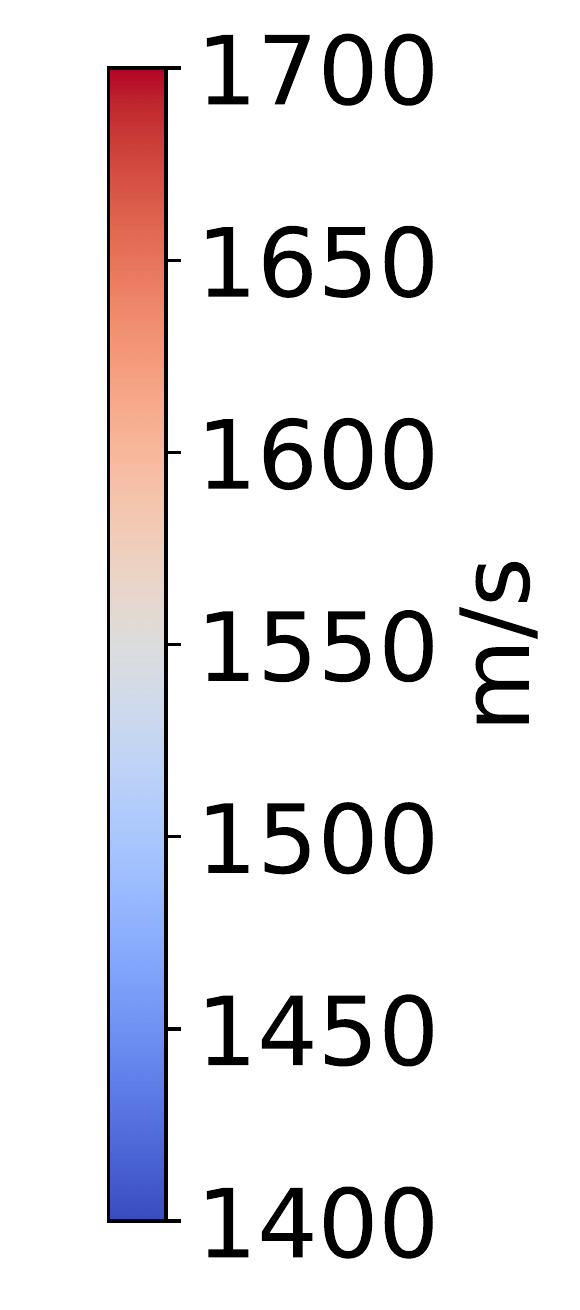}}\\
\raisebox{-.5\totalheight}{ \small \rot{\makecell{{En-De-Net} \\ {{SoS, B-mode}} \\ {Overlay} }}} & 
\raisebox{-.5\totalheight}{\includegraphics[  width =1.9cm]{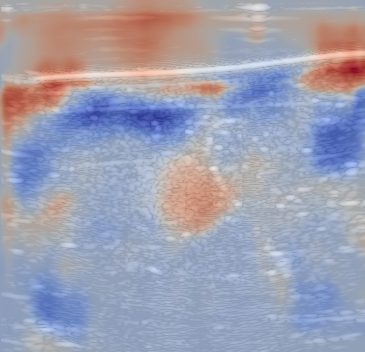}}&
\raisebox{-.5\totalheight}{\includegraphics[  width =1.9cm]{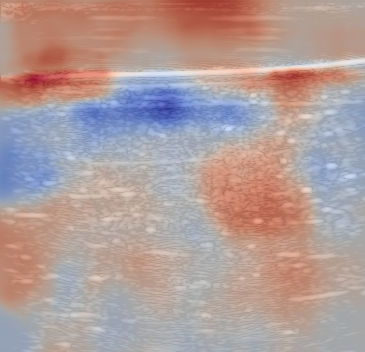}}&
\raisebox{-.5\totalheight}{\includegraphics[  width =1.9cm]{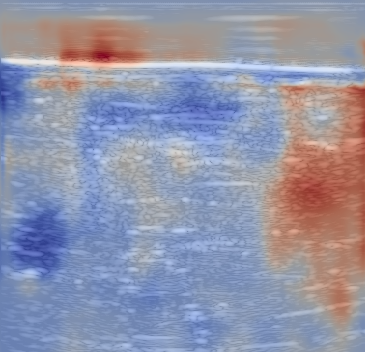}}&
\raisebox{-.5\totalheight}{\includegraphics[  width =1.9cm]{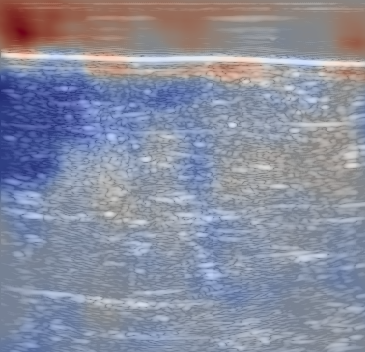}}&
\raisebox{-.5\totalheight}{\includegraphics[  width =1.9cm]{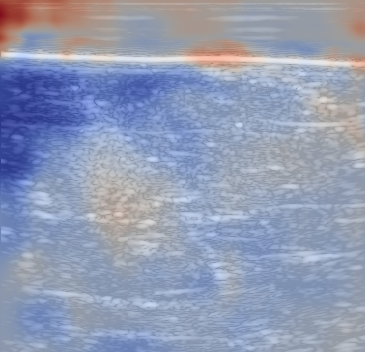}}&
\raisebox{-.5\totalheight}{\includegraphics[  width =1.9cm]{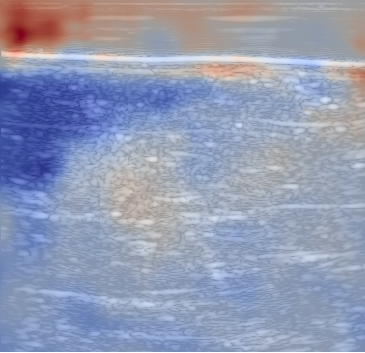}}&
\raisebox{-.5\totalheight}{\includegraphics[  width =1.9cm]{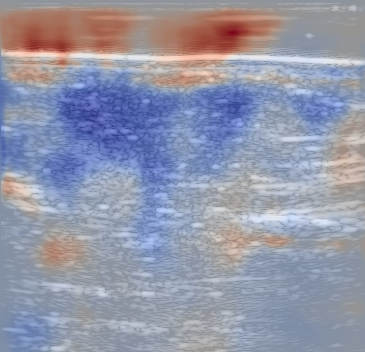}}&
\raisebox{-.5\totalheight}{\includegraphics[  width =1.9cm]{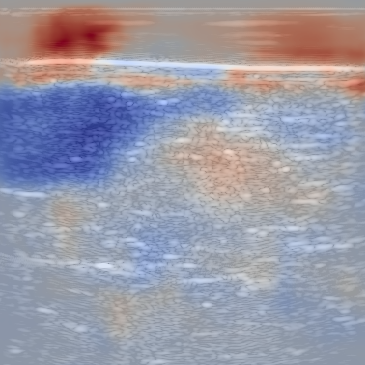}}&
\raisebox{-.5\totalheight}{\includegraphics[height =1.9cm]{img/colorbar_SoS_real.pdf}}\\

 \raisebox{-.5\totalheight}{ \small \rot{\makecell{{AutoSpeed} \\ {{SoS Map}} }}} & 
 \raisebox{-.5\totalheight}{\includegraphics[  width =1.9cm]{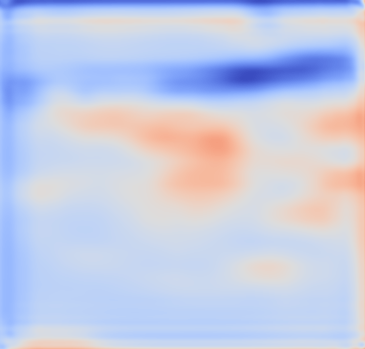}}&
 \raisebox{-.5\totalheight}{\includegraphics[  width =1.9cm]{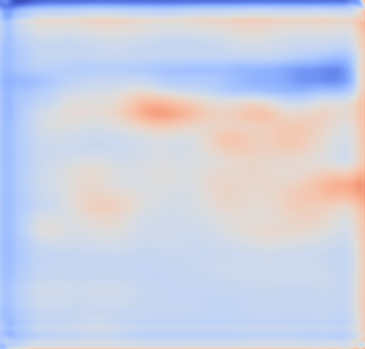}}&
 \raisebox{-.5\totalheight}{\includegraphics[  width =1.9cm]{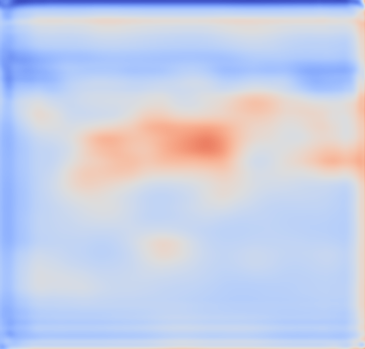}}&
 \raisebox{-.5\totalheight}{\includegraphics[  width =1.9cm]{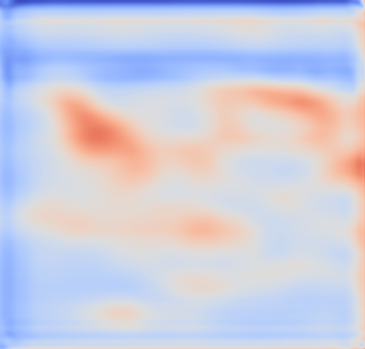}}&
 \raisebox{-.5\totalheight}{\includegraphics[  width =1.9cm]{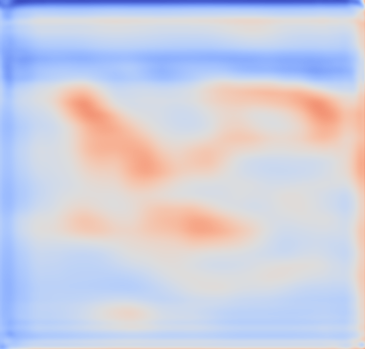}}&
 \raisebox{-.5\totalheight}{\includegraphics[  width =1.9cm]{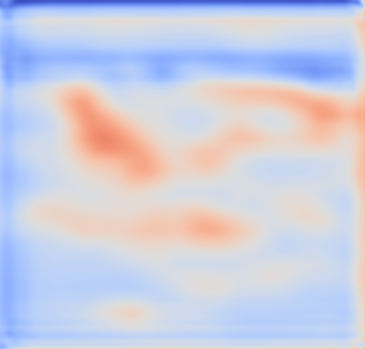}}&
 \raisebox{-.5\totalheight}{\includegraphics[  width =1.9cm]{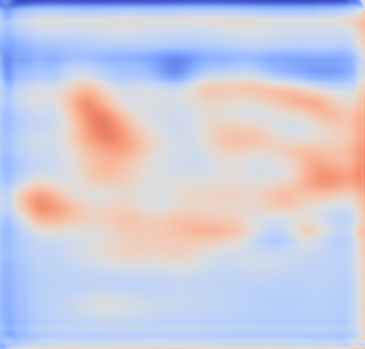}}&
 \raisebox{-.5\totalheight}{\includegraphics[  width =1.9cm]{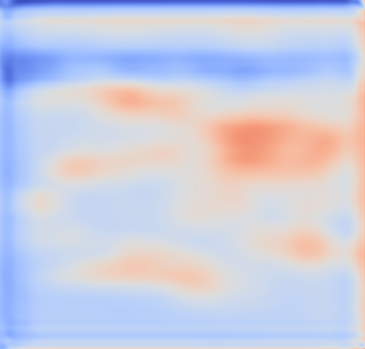}}&
 \raisebox{-.5\totalheight}{\includegraphics[height =1.9cm]{img/colorbar_SoS_real.pdf}}\\

\raisebox{-.5\totalheight}{ \small \rot{\makecell{{AutoSpeed} \\ {{SoS, B-mode}} \\ {Overlay} }}} & 
\raisebox{-.5\totalheight}{\includegraphics[  width =1.9cm]{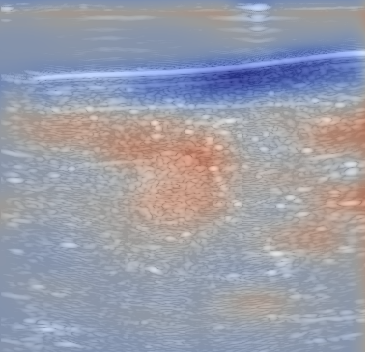}}&
\raisebox{-.5\totalheight}{\includegraphics[  width =1.9cm]{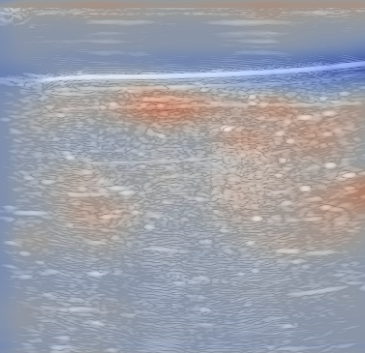}}&
\raisebox{-.5\totalheight}{\includegraphics[  width =1.9cm]{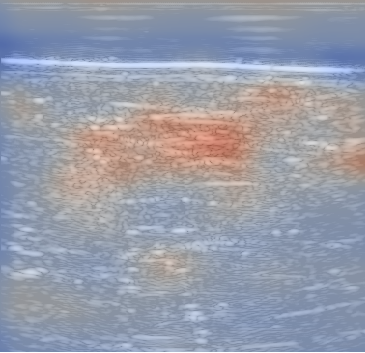}}&
\raisebox{-.5\totalheight}{\includegraphics[  width =1.9cm]{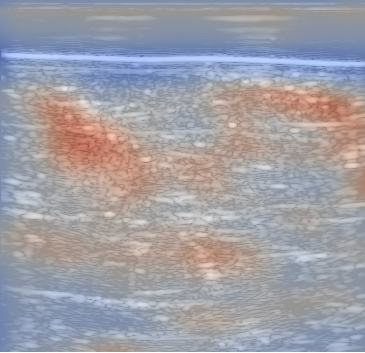}}&
\raisebox{-.5\totalheight}{\includegraphics[  width =1.9cm]{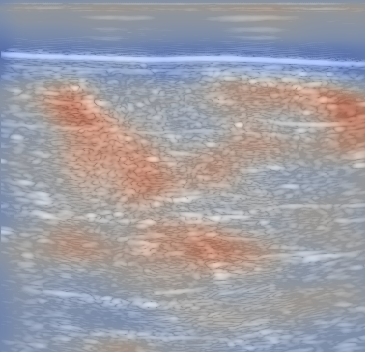}}&
\raisebox{-.5\totalheight}{\includegraphics[  width =1.9cm]{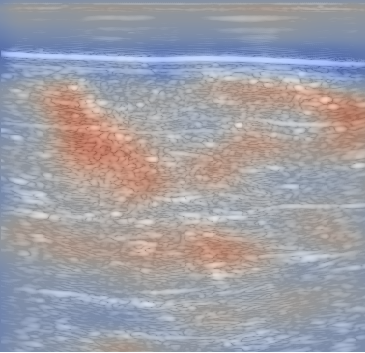}}&
\raisebox{-.5\totalheight}{\includegraphics[  width =1.9cm]{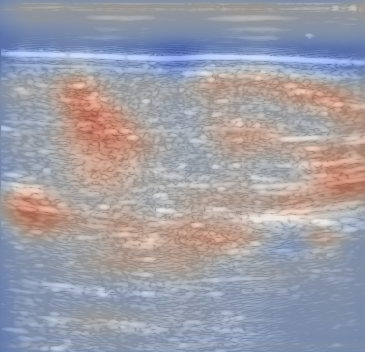}}&
\raisebox{-.5\totalheight}{\includegraphics[  width =1.9cm]{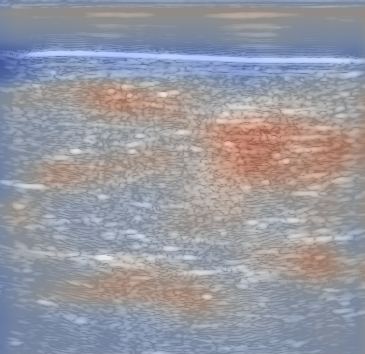}}&
\raisebox{-.5\totalheight}{\includegraphics[height =1.9cm]{img/colorbar_SoS_real.pdf}}\\

\end{tabular}
\caption{Predicted SoS maps by En-De-Net and AutoSpeed for 8 cases where a dense inclusion is present in the B-mode image. The inclusion is marked by red arrows in B-mode images.}
\label{fig: measured}
\end{figure*}

Fig.~\ref{fig: boxplot} shows predicted SoS values inside the inclusion and in the background for 30 cases compared to the expected value. For 30 cases, the mean predicted SoS value in the background is $1535$~$m/s$ with SD of $6$~$m/s$ and $1527$~$m/s$ with SD of $19$~$m/s$ for AutoSpeed and En-De-Net, respectively. 
The mean value of predicted SoS inside the inclusion for AutoSpeed is $1561$~$m/s$ with SD $11$~$m/s$ and for En-De-Net is $1545$~$m/s$ with SD of $43$~$m/s$ (Table~\ref{table: errors}). 
Fig.~\ref{fig: measured} shows 8 sample cases of this dataset. 
Quantitatively, the mean SoS values inside the inclusion and in the background are close to the expected values (Table~\ref{table: errors}). 
However, En-De-Net shows higher SD. The high SD for En-De-Net means that this network in some cases suffers from inverted SoS values, fails to find the inclusion and/or the predicted SoS values for the same kind of inclusion varies among shots. 
This effect can be seen in Fig.~\ref{fig: measured}, for example, in cases \#4, \#5, \#6, and \#7 which are very similar cases, En-De-Net shows varying SoS values in the inclusion region.  

\begin{figure}[!t]
     \begin{minipage}[b]{0.99\textwidth}
        \centering
        \captionof{table}{Mean SoS values in the background and inside dense inclusion acquired from CIRS multi-modality breast phantom, $\pm$ shows the SD value. The predicted values are computed on $30$ cases, all with inclusion in different locations.}
        \label{table: errors}
        \begin{tabular}{l|c|c}\hline
            & Background ($m/s$) & Inclusion ($m/s$) \\ \hline
            Expected value &  \(1520\pm10\) & \(1580\pm20\)  \\
            AutoSpeed & \(1535\pm 6\) & \(1561\pm11\)\\
            En-De-Net & \(1527\pm19\)  & \(1545\pm45\) \\ \hline
        \end{tabular}
    \end{minipage}
\end{figure}

Qualitatively, both networks in most cases can find the inclusion area with SoS contrast ($76\%$ and $90\%$ for En-Co-Net and AutoSpeed, respectively, for 30 cases).
However, if En-De-Net finds the area with SoS contrast often the predictions have clearer margins.
AutoSpeed shows more stable SoS predictions inside the inclusion area but in some cases shows SoS contrast in regions outside of the inclusion area, we noticed that these regions are the regions with high reflections compared to the background, for example, where highly reflective small structures are present.
Therefore, AutoSpeed is sensitive to these kinds of reflective scatterers. 
On the other hand, it is more likely that the predictions from one shot to another shot vary for the En-De-Net, meaning the predicted values inside the same inclusion contrary to expectation changes in different locations, e.g., Fig.~\ref{fig: measured} case \#2 vs. case \#3 or even in the very similar filed of views e.g., Fig.~\ref{fig: measured} case \#5 and \#5 vs. case \#7.
The following section shows the extreme case of this phenomenon, where multiple shots of the exact field of view with mechanically fixed transducer and phantom are investigated.

\textbf{Multiple Frames:} 
Here, we perform an experiment to demonstrate the stability and reproducibility of AutoSpeed predictions compared to En-De-Net when tested on multiple shots of the same field of view. 
We acquired 200 frames from the same field of view when the probe head and phantom is mechanically fixed. 
Since the field of view is the same, the expectation is that the networks predict consistent SoS values inside the inclusion and in the background. 
Additionally, we expect that the SoS values inside the inclusion have low SD. 
Fig.~\ref{fig: Sos Predictions, Auto vs EnDE} shows 4 cases of the aforementioned dataset. 

Although both networks follow the geometry in the corresponding b-mode images, could find the inclusion and the skin layer and background margin correctly, AutoSpeed shows more consistency. 
In Fig.~\ref{fig: Sos Predictions, Auto vs EnDE}, Frame 1 green arrows in the En-De-Net prediction image show the inconsistent regions that appear differently from frame to frame. 
For example, on the left side of the image, the SoS values are inverted from one frame to another frame. On the right side of the image, there is a region detected with SoS contrast, but the margin and the SoS values in the corresponding region vary. 
The same phenomenon can be seen in the inclusion area in the middle. 
Whereas, the predictions of the AutoSpeed network are consistent throughout all the frames. 

\begin{figure}[!t]
 \centering
    \renewcommand{\arraystretch}{0.1}
        \begin{tabular}{@{\hspace{0.5mm}} c  @{\hspace{0.5mm}} c @{\hspace{0.5mm}}c @{\hspace{0.5mm}}c @{\hspace{0.5mm}}c @{\hspace{0.5mm}}c }
        &  { Frame  1} &  { Frame  2} &  {Frame  3} &   {Frame  4} &  \\
        & & & & & \\ 
         \raisebox{-.5\totalheight}{ \small \rot {B-mode}} & 
         \raisebox{-.5\totalheight}{\includegraphics[    width = 2.3cm]{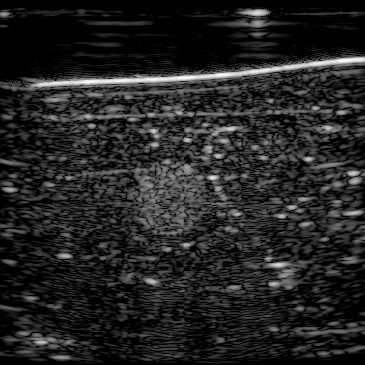}}&
         \raisebox{-.5\totalheight}{\includegraphics[    width = 2.3cm]{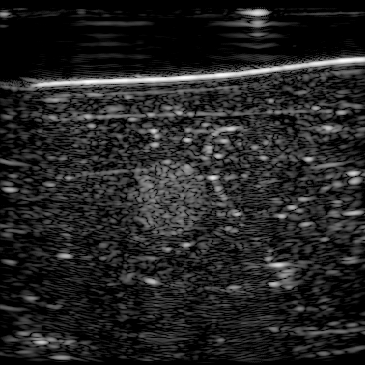}}&
         \raisebox{-.5\totalheight}{\includegraphics[    width = 2.3cm]{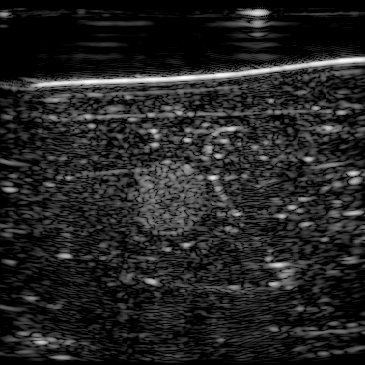}}&
         \raisebox{-.5\totalheight}{\includegraphics[    width = 2.3cm]{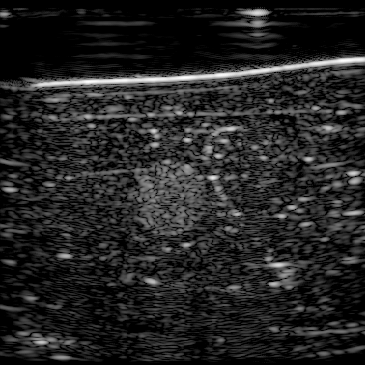}}&
        \raisebox{-.5\totalheight}{\includegraphics[height =2.3cm]{img/colorbar_bmode.pdf}}\\
        \\
        
        \raisebox{-.5\totalheight}{ \small \rot{\makecell{{Absolute} \\  {Difference} \\ {B-mode}}}}& 
         \raisebox{-.5\totalheight}{\includegraphics[    width = 2.3cm]{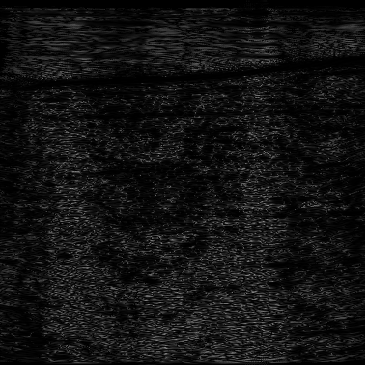}}&
         \raisebox{-.5\totalheight}{\includegraphics[    width = 2.3cm]{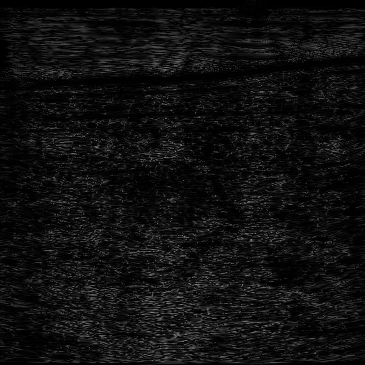}}&
         \raisebox{-.5\totalheight}{\includegraphics[    width = 2.3cm]{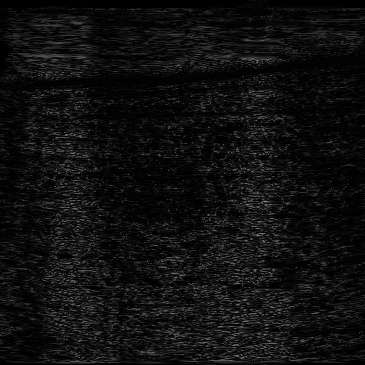}}&
         \raisebox{-.5\totalheight}{\includegraphics[    width = 2.3cm]{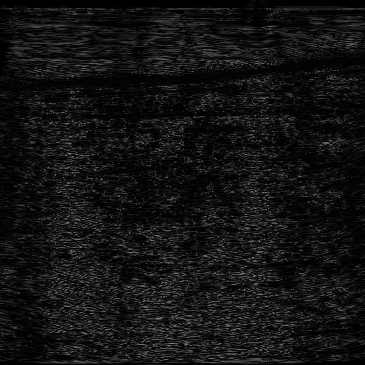}}&
        \raisebox{-.5\totalheight}{\includegraphics[height =2.3cm]{img/colorbar_bmode.pdf}}\\
        \\
        
         \raisebox{-.5\totalheight}{ \small \rot{\makecell{ {En-De-Net} \\{SoS Map} }}}& 
         \raisebox{-.5\totalheight}{\includegraphics[    width = 2.3cm]{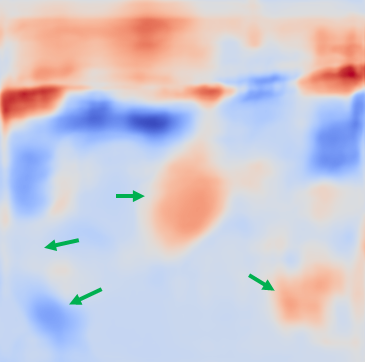}}&
         \raisebox{-.5\totalheight}{\includegraphics[    width = 2.3cm]{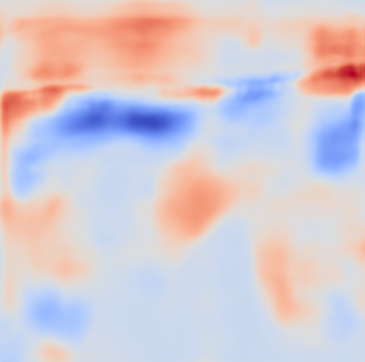}}&
         \raisebox{-.5\totalheight}{\includegraphics[    width = 2.3cm]{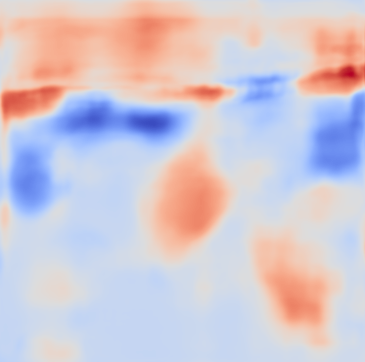}}&
         \raisebox{-.5\totalheight}{\includegraphics[    width = 2.3cm]{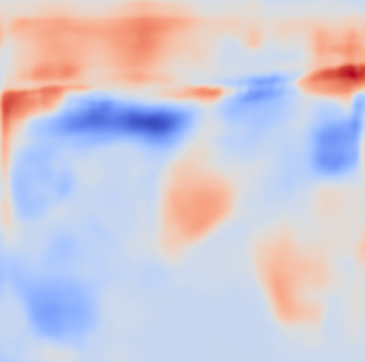}}&
        \raisebox{-.5\totalheight}{\includegraphics[height =2.3cm]{img/colorbar_SoS_real.pdf}}\\
        \\
        
         \raisebox{-.5\totalheight}{ \small \rot{\makecell{{AutoSpeed} \\ {SoS Map}}}}& 
         \raisebox{-.5\totalheight}{\includegraphics[    width = 2.3cm]{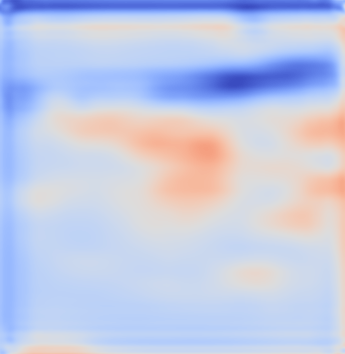}}&
         \raisebox{-.5\totalheight}{\includegraphics[    width = 2.3cm]{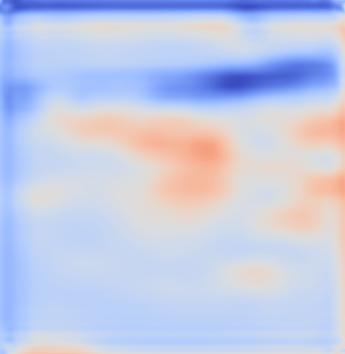}}&
         \raisebox{-.5\totalheight}{\includegraphics[    width = 2.3cm]{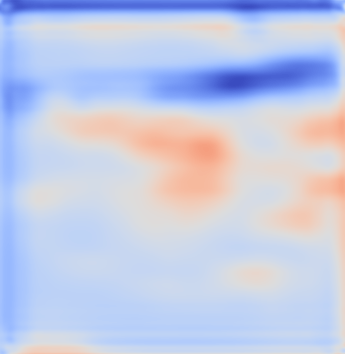}}&
         \raisebox{-.5\totalheight}{\includegraphics[    width = 2.3cm]{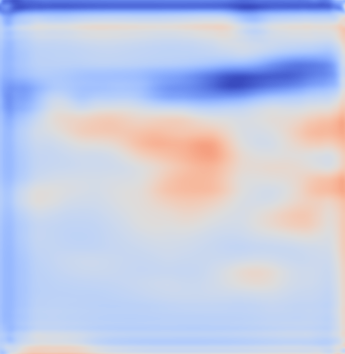}}&
        \raisebox{-.5\totalheight}{\includegraphics[height =2.3cm]{img/colorbar_SoS_real.pdf}}\\
        \\
        
        \end{tabular}
        \caption{1st row: reconstructed b-mode images of 4 consecutive frames, 2nd row: absolute difference matrices of each frame with its previous frame (frame 0 is not shown here), 3rd row: predicted SoS for the consecutive shots for En-De-Net, 4th row: predicted SoS for AutoSpeed Network. Predicted SoS maps by AutoSpeed are more consistent for the same field of view compared to the predictions of En-De-Net.}
        \label{fig: Sos Predictions, Auto vs EnDE}
\end{figure}

\begin{figure}[!t]
\centering 
\renewcommand{\arraystretch}{0.05}
\begin{tabular}{ @{\hspace{0.5mm}}c | @{\hspace{0.5mm}}c   }

 Inside Inclusion & Background \\

\raisebox{-.5\totalheight}{\includegraphics[width=6.5cm]{ 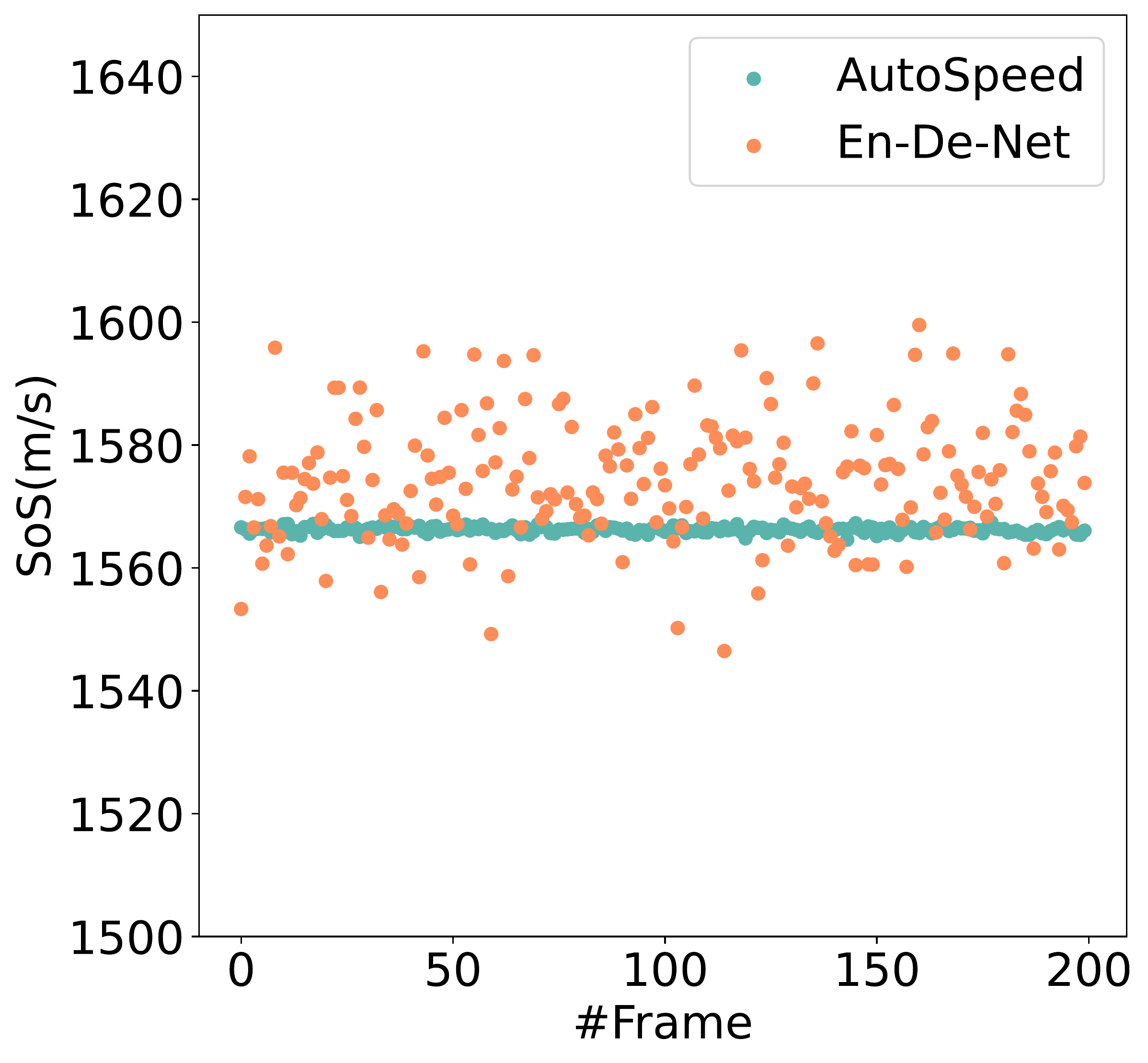}}&
\raisebox{-.5\totalheight}{\includegraphics[width=6.5cm]{ 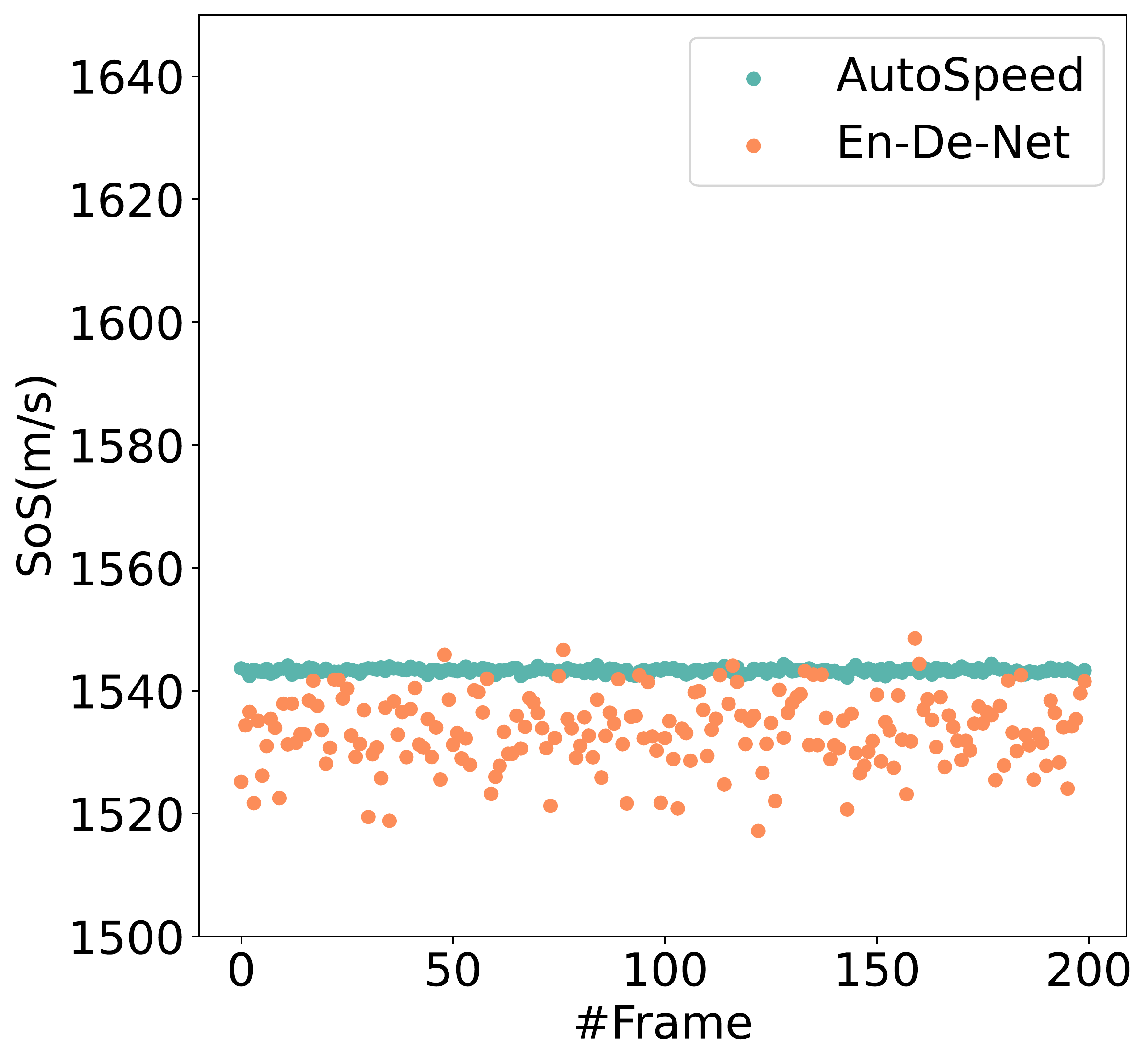}}\\
(a) & (b) \\ 
\raisebox{-.5\totalheight}{\includegraphics[width=6.5cm]{ 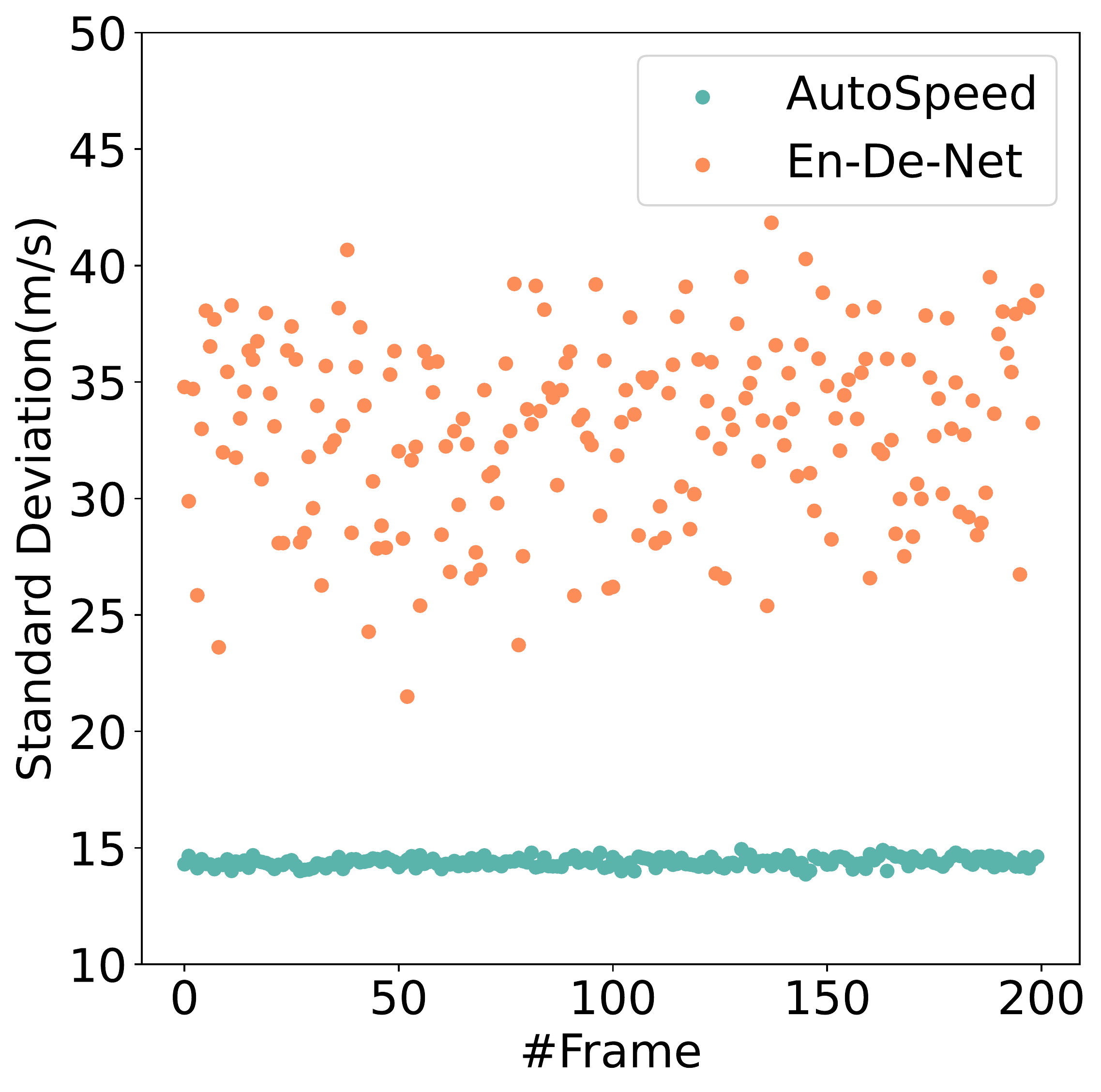}}&
\raisebox{-.5\totalheight}{\includegraphics[width=6.5cm]{ 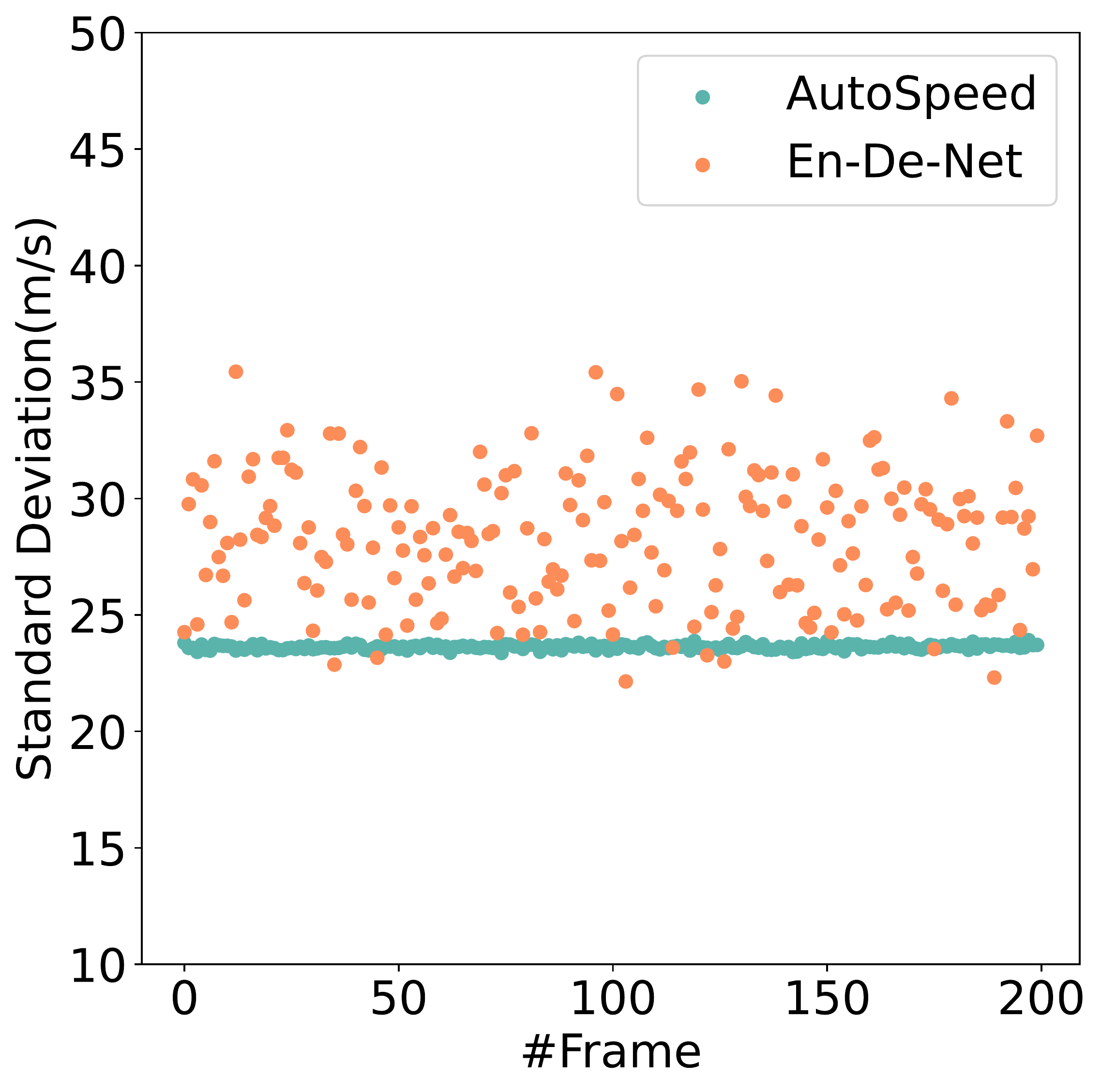}}\\
(c) & (d) \\ 
\end{tabular}
 \caption{AutoSpeed vs. En-De-Net, over 200 consecutive frames from the same field of view: \textbf{(a)}: Comparison of average predicted SoS values inside the inclusion, \textbf{(b)}: Comparison of average predicted SoS values in the background,  \textbf{(c)}: Comparison of the standard deviation of the predictions inside the inclusion, \textbf{(d)}:  Comparison of the standard deviation of the predictions in the background; average predicted SoS values by AutoSpeed both inside the inclusion and in the background is more consistent through all the frames compared to En-De-Net. }
\label{fig: scatterplot, frames, Auto vs EnDe}
\end{figure}

Fig.~\ref{fig: scatterplot, frames, Auto vs EnDe} compares the mean SoS values inside the inclusion and in the background region for two networks and the standard deviations of predicted values inside the inclusion area and in the background for each frame. 
Based on Fig.~\ref{fig: scatterplot, frames, Auto vs EnDe} (a) and (b), AutoSpeed has highly consistent predictions both inside the inclusion and in the background in consecutive frames of the same field of view. 
Additionally, Fig.~\ref{fig: scatterplot, frames, Auto vs EnDe} (c) and (d) show that the SoS values in each frame inside the inclusion and in the background have consistent uniformity for AutoSpeed compared to En-De-Net. 
Thus, we can conclude that AutoSpeed can extract more stable features compared to En-De-Net which performs an end-to-end mapping and lower error rates on the simulated data do not necessarily translate to the measured data.

\section{Discussion and Conclusion}

SoS reconstruction for pulse-echo ultrasound can be advantageous in clinical studies, because it employs the available data acquisition setups used for B-mode imaging and it provides quantitative measures that can be used to distinguish different tissue types. 
Yet, it is still in its infancy stage and there is no accepted gold standard approach for pulse-echo SoS reconstruction. 
Analytical and optimization-based methods require prior and carefully chosen regularization parameters and the proposed methods are often either not stable and/or not feasible for real-time applications.
In recent years, deep learning techniques are widely used to solve inverse problems. 
These methods often perform an par or even in cases outperform model-based methods. 
Additionally, they move the computational burden to training phase, thus, in inference, they can perform in real-time. 
But they need training data often alongside their labels or GT maps. 
Hence, the lack of a gold standard method for SoS reconstruction poses a challenge to develop deep learning techniques as well. 
Because it is practically impossible to create sufficient GT data for deep learning algorithms.
Simulation toolboxes offer an alternative solution to create training data.
However, there is no guarantee that the networks trained on simulated data are transferable to real data. 
As a result, incorporating prior knowledge can increase the chance of success of such networks in real data setups. 

In this study, we proposed a novel approach for pulse-echo SoS imaging from a single plane-wave acquisition.
Inspired by the known parameter paradigm \cite{maier2019learning,maier2022known}, we hardcoded the domain transfer problem in a dual autoencoder approach, similar to the idea behind Cycle-GAN \cite{zhu2017unpaired}. 
We trained two linked autoencoders to extract efficient representations from RF data and SoS maps. In a fine-tuning step, a fully connected layer mapped the representation from the RF data domain to the SoS domain. 
There is evidence that the functions learned by networks can be decomposed into modules and those modules can be used in other tasks even without any further training \cite{maier2022known,fu2019divide,fu2020modularization}. Thus, in the inference, an encoder-decoder-like architecture is proposed in which the encoder is detached from the RF data autoencoder and the decoder is detached from the SoS autoencoder (both fully trained) and two paths are connected via  a trained fully connected layer (IRM-Layer). 
We tested the method on simulated and measured data and made a comparison with an encoder-decoder network  that is trained end-to-end. 
We showed that SoS mapping is possible employing such a setup and is more stable compared to the end-to-end mapping solution previously proposed. 

On the simulated data, the proposed network, i.e., AutoSpeed, reconstructs the SoS maps with $37.26\pm3.56$~$m/s$ MAE is in the typical range of SoS error rates but the error rates are higher compared to the network that is trained end-to-end, i.e., En-De-Net which is $17.32\pm0.7$~$m/s$.
\cite{li2009vivo} demonstrated that breast fatty tissue and parenchyma have mean SoS values of $1422\pm9$~$m/s$ and $1487\pm21$~$m/s$, respectively. The SoS values for lesions is on average higher, e.g., benign lesions have a mean SoS value of $1513\pm27$~$m/s$ and for malignant lesions, the average SoS is $1548\pm17$~$m/s$.
Another study \cite{ruby2019breast} showed that the presence of breast lesions results in focal increments $\Delta$SoS in comparison with the background tissue. 
Malignant lesions show significantly higher $\Delta$SoS in comparison with begin lesoins $\Delta SoS > 41.64 $~$m/s$.
\cite{ruby2019breast} reported $\Delta$SoS in range $[14-118]$~$m/s$ for malignant and $[7-41]$~$m/s$ for benign lesions. 
Therefore, theoretically, the predicted values are relevant for clinical use cases.

Our experiments on the measured data demonstrated that outperforming on the simulated data does not necessarily translate to the measured data setup and the networks that are trained end-to-end can be prone to overfitting to the distribution of simulated data.
Whereas, our proposed method that employs the learned representations from autoencoders is more stable, can extract efficient features, and outperforms the previously proposed method in terms of reproducibility.
On the measured data setup, AutoSpeed could detect the inclusions and correct margins, and the predicted SoS maps are close to the expected range.

Initial results on phantom data are highly encouraging. 
AutoSpeed on the measured data predicted SoS values in range $1535\pm6$~$m/s$ and $1561\pm11$~$m/s$ in the background region and inside the inclusion area, respectively, where the corresponding expected values are $1520\pm10$~$m/s$ and $1580\pm20$~$m/s$. 
Nevertheless, further research is required to improve the results, e.g., remove artifacts related to highly reflective scatterers, for instance, by including similar structures in the tissue modeling, and improve training to create cleaner margins, for example, by using multi-scale approaches. 
Additionally, more research is required to transfer and prove the efficiency of such methods in clinical setups.

\section*{Disclaimer}
The information in this paper is based on research results that
are not commercially available.

\section*{Acknowledgments}
We thank Prof. Michael Golatta and the university hospital of Heidelberg for providing the Tomosynthesis dataset.

\bibliographystyle{unsrt}  

\bibliography{refs}

\end{document}